\documentclass[12pt,superscriptaddress]{article}
\usepackage{palatino,cite,epsfig,amsmath,amssymb}

\allowdisplaybreaks

\begin{document}
\begin{titlepage}

\begin{flushright}
Edinburgh 2004/06\\
SHEP-04-08\vspace{0.5cm}\\
\end{flushright}

\vspace{1cm}

\begin{center}
{\Large \bf An Interesting NMSSM Scenario at the LHC and LC:\\[0.3cm]
\large A Contribution to the LHC / LC Study Group}\\[1cm]
{\large D.J.~Miller$^{1}$ and S.~Moretti$^{2}$}\\[1cm]
{\it $^1$ School of Physics, The University of Edinburgh, Edinburgh EH93 JZ, Scotland\\
     $^2$ School of Physics and Astronomy, The University of Southampton, \\
Highfield, Southampton S017 1BJ, UK}\\
\end{center}

\renewcommand{\thefootnote}{\fnsymbol{footnote}}
\vspace{3cm}

\begin{abstract}
\noindent
The Next--to--Minimal Supersymmetric Standard Model NMSSM provides an
attractive extension to the minimal supersymmetric model by including
an extra Higgs singlet superfield. This extension allows one to link
the Higgs-higgsino mass parameter $\mu$ to a vacuum expectation value
of the new scalar field, thus providing a solution to the
$\mu$--problem of the MSSM. It this report, presented within the
context of the \linebreak[4] LHC / LC Study Group, we examine a
particularly interesting NMSSM scenario where the extra Higgs scalar
is rather light. We determine LHC production cross-sections and
branching ratios for the lightest scalar and find that it will be
difficult to observe at the LHC. However, we show that this lightest
scalar can instead be observed at an $e^+e^-$ Linear Collider for all
but a small window of parameter space.
\end{abstract}

\end{titlepage}

\subsection*{1~~~Introduction}

The Next-to-Minimal Supersymmetric Standard Model (NMSSM) provides an
elegant solution to the $\mu$ problem of the MSSM by introducing an
extra complex scalar Higgs superfield. The extra fields have no gauge
couplings and are principally only manifest through their mixing with
the other states. This leads to scenarios where Higgs boson couplings
are reduced in comparison to the MSSM, presenting a challenge to the
next generation of colliders. It is important that these extra Higgs
states be seen in addition to the expected Higgs doublet states in
order to distinguish the NMSSM from the MSSM. In this contribution, we
will examine the phenomenology of one of these scenarios at the LHC
and a future $e^+e^-$ Linear Collider and demonstrate a synergy
between the two machines.

The NMSSM has already been discussed in Section 2.4.1 of this study,
in the context of establishing a ``no-lose'' theorem for the discovery
of at least one Higgs boson at the next generation of colliders (see
also Ref.\cite{Ellwanger:2003jt}). It was seen that for some
exceptional NMSSM parameter choices the discovery of {\it any} Higgs
boson at all will be difficult at the LHC, but for the majority of
choices at least one Higgs boson will be discovered. Here we adopt a
different philosophy and examine a ``typical'' NMSSM scenario
point. While not representative of scenarios over the entire range of
parameters, the chosen scenario is certainly not unusual and a wide
range of parameter choices will result in similar phenomenology,
differing only in numerical detail and not in general structure. This
scenario therefore presents an interesting illustrative picture of the
Higgs sector that might be waiting to be explored in its full complexity
at the next generation of colliders.

\subsection*{2~~~The Model}

The NMSSM has the same field content as the minimal model augmented by
an additional neutral singlet superfield $\hat{S}$. Its superpotential
is given by
\begin{equation}
W=\hat{u}^c \, \mathbf{h_u} \hat{Q} \hat{H}_u  
-\hat{d}^c \, \mathbf{h_d} \hat{Q} \hat{H}_d  
-\hat{e}^c \, \mathbf{h_e} \hat{L} \hat{H}_d  
+\lambda \hat{S}(\hat{H}_u \hat{H}_d)+\frac{1}{3}\kappa\hat{S}^3,
\label{eq:superpotential} 
\end{equation}
where $\hat{H}_u$ and $\hat{H}_d$ are the usual Higgs doublet
superfields with $\hat{H}_u\hat{H}_d \equiv \hat{H}_u^+\hat{H}_d^- -
\hat{H}_u^0\hat{H}_d^0$. $\hat{Q}$ and $\hat{L}$ represent left
handed quark and lepton weak isospin doublets respectively, while
$\hat{u}^c$, $\hat{d}^c$ and $\hat{e}^c$ are the right handed quark
and lepton fields; $\mathbf{h_u}$, $\mathbf{h_d}$ and $\mathbf{h_e}$
are matrices of Yukawa couplings where family indices have been
suppressed.  The usual $\mu$-term of the MSSM, $\mu \hat{H}_u
\hat{H}_d$, has been replaced by a term coupling the new singlet field
to the usual Higgs doublets, $\lambda \hat{S} \hat{H}_u
\hat{H}_d$. When the new singlet field gains a vacuum expectation
value (VEV), an effective $\mu$-term is generated with an effective
Higgs-higgsino mass parameter given by $\mu_{\rm eff}=\lambda \langle
S \rangle$. (We adopt the notation that the superfields are denoted by
expressions with a ``hat'', while their scalar components are denoted
by the same expression without the hat.) The superpotential resulting
from this extension of the minimal model still contains an extra
symmetry [before the kappa term in Eq.(\ref{eq:superpotential}) is
included] --- a $U(1)$ ``Peccei-Quinn'' (PQ)
symmetry~\cite{Peccei:1977hh}, which will be broken when the singlet
field gains a non-zero VEV.  This spontaneous breaking results in a
massless Nambu-Goldstone boson which is in this instance a
pseudoscalar Higgs state. [In contrast the corresponding lightest
scalar Higgs state is {\it not} massless.]  Since this Higgs state has
not been observed in experiment we have only two possibilities: we
must either break the Peccei-Quinn symmetry explicitly, giving the
pseudoscalar a mass and putting it out of the kinematical reach of
past experiments, or we must decouple it from the other particles by
setting $\lambda \ll 1$. Here we adopt the former
possibility\footnote{For a description of the decoupled case, see
Ref.\cite{Miller:2003hm}.}, introducing an {\it explicit} Peccei-Quinn
symmetry breaking term $\frac{1}{3}\kappa\hat{S}^3$. This results in
the superpotential given in Eq.(\ref{eq:superpotential}). We will not
elaborate on the formal details of the model here except to elucidate
our parameter choice --- for a more detailed examination of the model
see Ref.\cite{Miller:2003ay} and references therein.

At tree level, the NMSSM Higgs sector has seven parameters: the Higgs
couplings from the superpotential, $\lambda$ and $\kappa$; their two
associated soft supersymmetry breaking parameters, $A_{\lambda}$ and
$A_{\kappa}$ ; and the VEVs of the three neutral Higgs fields, which
we re-express as two ratios of VEVs, $\tan \beta=\langle H_u^0 \rangle
/ \langle H^0_d \rangle$ and $\tan \beta_s=\sqrt{2} \langle S
\rangle/v$, and the electroweak scale $v/\sqrt{2}=\sqrt{\langle H^0_u
\rangle^2+\langle H^0_d \rangle^2}$.  The scenario to be considered
here has parameters given by $\lambda=0.3$, $\kappa=0.1$, $\tan \beta
= \tan \beta_s =3$ and $A_{\kappa}=-60$~GeV. The parameter
$A_{\lambda}$ is replaced by the mass scale $M_A$ which is chosen to
be the diagonal entry of the pseudoscalar Higgs boson mass-squared
matrix that returns to the value of the physical MSSM pseudoscalar
Higgs boson mass in the MSSM limit (i.e.\ $\lambda \to 0$, $\kappa \to
0$ while keeping $\lambda/\kappa$ and $\mu_{\rm eff}$ fixed). This
choice allows the reader a more intuitive connection with the
MSSM. $M_A$ will not be fixed, but will be allowed to vary over the
physical range. Finally, we take $v=246$~GeV.

\subsection*{3~~~The Mass Spectrum}

\begin{figure}[htb]
\begin{center}
\includegraphics[scale=0.6]{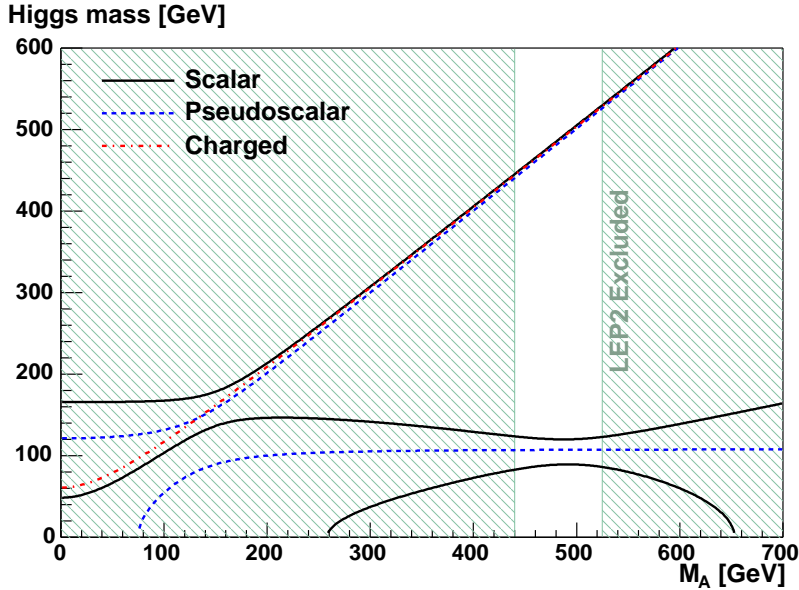}
\caption{\it 
The one-loop Higgs mass spectrum as a function of $M_A$ for
$\lambda=0.3$, $\kappa=0.1$, $\tan \beta= \tan \beta_s =3$ and
$A_{\kappa}=-60$~GeV. Also shown by the shaded area are the values of
$M_A$ that are ruled out by LEP2~\cite{Barate:2003sz} for this
parameter choice.}
\label{fig:masses}
\end{center}
\end{figure}

The Higgs mass spectrum for our parameter choice, evaluated at
one-loop precision~\cite{Kovalenko:dc}, can be seen in Fig.(\ref{fig:masses}),
as a function of $M_A$. This spectrum looks remarkably like that of
the MSSM with the addition of two extra Higgs fields --- a scalar
state and a pseudoscalar state. As in the MSSM, the heavy
pseudoscalar, scalar and charged Higgs bosons all lie around the mass
scale $M_A$, while a lighter scalar state has mass around
$115$-$130$~GeV. However, in addition we see extra scalar and
pseudoscalar states with masses of order $100$~GeV and below; these
are the Higgs states which are dominated by the extra singlet degrees
of freedom.

Making an expansion in the (often) small parameters $1/\tan \beta$ and
$M_Z/M_A$ allows us to obtain simple approximate forms for the masses
of these extra singlet dominated Higgs
bosons~\cite{Miller:2003hm}. One finds that the singlet dominated
pseudoscalar Higgs fields has a mass given approximately by
\begin{equation}
M_{A_1}^2 \approx -\frac{3}{\sqrt{2}} \kappa v_s A_{\kappa}, \label{eq:mas}
\end{equation}
while the singlet dominated scalar has a mass which is maximized at 
$M_A \approx 2 \mu_{\rm eff}/\sin 2\beta$ where it is given by
\begin{equation}
M_{H_1}^2 \approx \frac{1}{2} \kappa v_s (4\kappa v_s+\sqrt{2}A_{\kappa}). \label{eq:mah}
\end{equation}
It must be stressed that these expressions are approximate and are not
applicable over the entire parameter range; the one-loop expressions
for the masses should be used in preference, as in
Fig.(\ref{fig:masses}). However, the approximate expressions are
useful in determining the qualitative behaviour of the masses as the
parameters are varied. [Although approximate, these expressions do
surprisingly well in estimating the singlet dominated masses. For
example, for the present parameter choice they give $M_{A_1} \approx
96.2$~GeV and $M_{H_1} \approx 88.1$~GeV, which compare favourably
with the one-loop results, $107.3$~GeV and $89.5$~GeV respectively at
$M_A=495$~GeV. This is in part due to the suppression of couplings to
quarks, which reduces the impact of radiative corrections.]

In particular, the masses are strongly dependent only on the
quantities $\kappa v_s$ and $A_{\kappa}$ [and $M_A$].  The dependence
on $\kappa v_s$ (which is a measure of how strongly the PQ symmetry is
broken) is straightforward: as $\kappa v_s$ increased the masses also
increase.  Since one expects $v_s$ to be of the order of $v$ and
$\kappa$ is restricted by $\kappa^2+\lambda^2 \lesssim 0.5$ when one
insists on perturbativity up to the unification scale, it is natural
(though not mandatory) for this mass scale to be rather low, and the
extra Higgs states rather light. In contrast, the $A_{\kappa}$
contribution to the masses has opposite sign for scalar and
pseudoscalar. The dependence of the pseudoscalar mass,
Eq.(\ref{eq:mas}), on $A_{\kappa}$ indicates that $A_{\kappa}$ should
be negative, while Eq.(\ref{eq:mah}) insists that its absolute value
does not become too large. These effects are nicely summarized by the
approximate mass sum rule (at $M_A \approx 2
\mu_{\rm eff}/\sin 2\beta$):
\begin{equation}
M_{H_1}^2+\frac{1}{3}M_{A_1}^2 \approx 2 \; (\kappa v_s)^2.  \label{eq:sumrule}
\end{equation}
The overall scale for the masses is set by $\kappa v_s$, while
increasing the scalar mass leads to a decrease in the pseudoscalar
mass and vice versa.

Fig.(\ref{fig:masses}) also shows the values of $M_A$ that, for this
parameter choice, are already ruled out by LEP (the shaded
region). Although a SM Higgs boson with mass below $114.4$~GeV is now
ruled out with $95\%$ confidence by the LEP
experiments~\cite{Barate:2003sz}, lighter Higgs bosons are still
allowed if their coupling to the $Z$ boson is reduced. In the NMSSM,
since the extra singlet fields have no gauge couplings, the couplings
of the singlet dominated fields to the $Z$ boson come about only
through mixing with the neutral doublet Higgs fields. When this mixing
is small their couplings are reduced and they can escape the
Higgs-strahlung dominated LEP limits. For the LEP limits shown here we
take into account decays to both $b \bar b$~\cite{Barate:2003sz} and
$\gamma \gamma$~\cite{lhwg-2002-02}, as well as decay mode
independent searches carried out by the OPAL
detector~\cite{Abbiendi:2002qp}. As expected, the limits are dominated
by the decay $H_1 \to b \bar b$.

The dependence of the lightest Higgs boson mass on $M_A$ also makes a
prediction for the mass of the heavy states. The lightest Higgs boson
mass must be kept large enough to escape the current LEP limits.
However, since this mass decreases rapidly to either side of its
maximum (see Fig.(\ref{fig:masses}) we are forced to constrain $M_A$,
and thus the heavy Higgs boson masses, to around $M_A \approx 2
\mu_{\rm eff}/\sin 2\beta \approx \mu_{\rm eff} \tan\beta$.

There is still significant room for a rather light Higgs
bosons to be found the LHC and/or a LC. It is essential that these
light Higgs bosons be ruled out or discovered at the next generation
of colliders. In the following we will focus on the production of a
light singlet dominated scalar Higgs boson at the LHC and a LC and its
subsequent decay, but one should bear in mind that there is also
a light pseudoscalar Higgs boson which also deserves study.

\subsection*{4~~~Branching ratios for the light scalar}

\begin{figure}[htb]
\begin{center}
\includegraphics[scale=0.6]{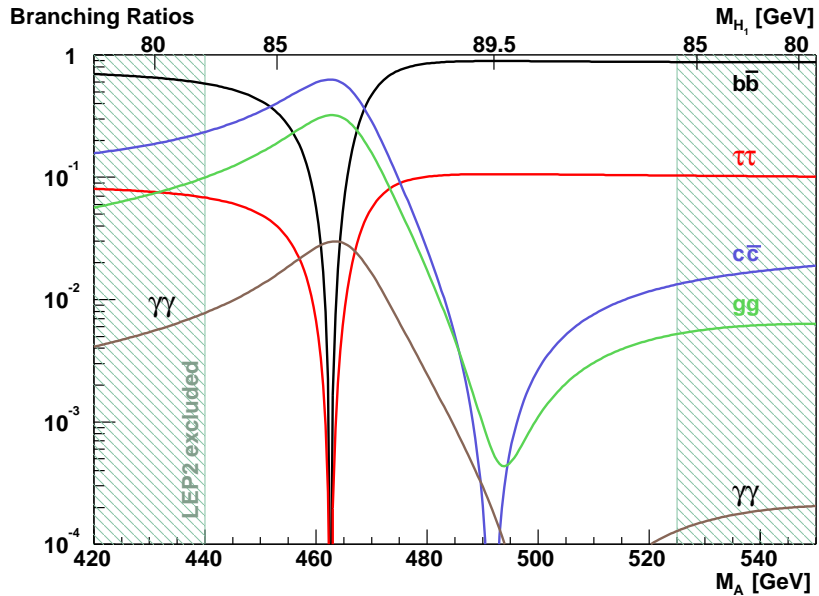}
\caption{\it 
The dominant branching ratios for the lightest scalar Higgs boson as a
function of $M_A$ for $\lambda=0.3$, $\kappa=0.1$, $\tan \beta= \tan
\beta_s =3$ and $A_{\kappa}=-60$~GeV. The complicated structure is due
to the switching off of the Higgs boson couplings to up-type and
down-type quarks and leptons.}
\label{fig:branchingratios}
\end{center}
\end{figure}

The dominant branching ratios of the lightest scalar Higgs boson are
shown in Fig.(\ref{fig:branchingratios})
as a function of $M_A$. For a SM Higgs boson of the same mass (around
$80-90$~GeV) one would expect the dominant decays to be to bottom
quarks, $\tau$ leptons, and charm quarks, with the addition of loop
induced decays to gluons and photons. These are indeed also the
dominant decays of the singlet dominated scalar for most of the
allowed $M_A$ range, but the branching ratios now show significant
structure at approximately $463$~GeV and again at around $490$~GeV due
to the suppression of various couplings.

The couplings of the lightest Higgs scalar to up-type and
down-type quarks and leptons are given in terms of the SM Higgs
couplings by
\begin{eqnarray}
g^{\rm NMSSM}_{H_1u \bar u} &=& 
(\phantom{-} O^H_{11} \cot \beta + O^H_{21})\; g^{\rm SM}_{Hu \bar u}, \label{eq:ghuu} \\
g^{\rm NMSSM}_{H_1d \bar d} &=& 
(- O^H_{11} \tan \beta + O^H_{21})\; g^{\rm SM}_{Hd \bar d}, \label{eq:ghdd} 
\end{eqnarray}
respectively, where $O^H_{11}$ and $O^H_{21}$ are elements of the
scalar Higgs mixing matrix. The relative minus sign between terms in
Eq.(\ref{eq:ghuu}) and Eq.(\ref{eq:ghdd}) has the same origin as the
relative minus sign between the $hu \bar u$ and $hd \bar d$ couplings
in the MSSM.

The first structure seen in Fig.(\ref{fig:branchingratios}), at around
$463$~GeV, is due to the cancellation of $-O^H_{11} \tan \beta$ with
$O^H_{21}$ in Eq.(\ref{eq:ghdd}), forcing the $H_1 \to b \bar b$ and
$H_1 \to \tau^+ \tau^-$ branching ratios to vanish. As $M_A$ is
increased, $O^H_{21}$ passes smoothly through zero, eventually
canceling with $O^H_{11} \cot \beta$ in Eq.(\ref{eq:ghuu}). This
provides the structure at around $490$~GeV where the $H_1 \to c \bar
c$ branching ratio vanishes. 

The decays to $gg$ and $\gamma \gamma$ are mediated by loop diagrams
giving a more complex behaviour. $H_1 \to gg$ is dominated by top and
stop loops and consequently shows a marked decrease as the $H_1 t \bar
t$ coupling switches off; although the top-loop contribution will pass
through zero here, stop loops and bottom (s)quark loops prevent the
branching ratio from vanishing. In addition to top and bottom (s)quark
loops the $\gamma \gamma$ branching ratio is mediated by virtual $W$
bosons, charged Higgs bosons and charginos. The dominant effect is
from the $W$ bosons and the top loops and so we see a broad
suppression over the range where these couplings vanish.

\subsection*{5~~~LHC Production}

\begin{figure}[htb]
\begin{center}
\includegraphics[scale=0.6]{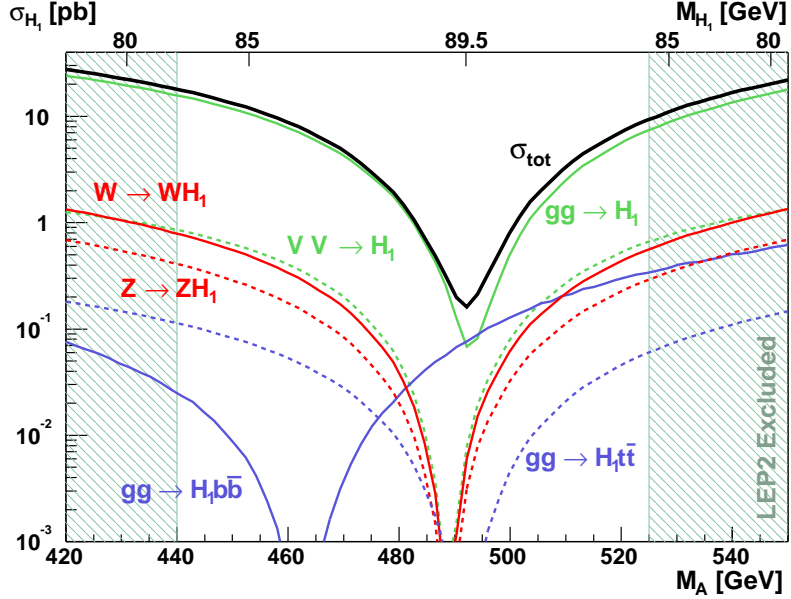}
\caption{\it 
Production cross-sections for the lightest scalar Higgs boson at the
LHC, as a function of $M_A$ for $\lambda=0.3$, $\kappa=0.1$, $\tan
\beta= \tan \beta_s =3$ and $A_{\kappa}=-60$~GeV. }
\label{fig:lhc}
\end{center}
\end{figure}

Cross-sections for the production of the lightest scalar Higgs boson
in various channels at the LHC are shown in Fig.(\ref{fig:lhc}).
The total production cross-section is dominated by gluon-gluon fusion,
and is sizable over the entire range. Other significant production
channels are vector boson fusion ($VV \to H_1$), Higgs-strahlung ($W
\to WH_1$ and $Z \to ZH_1$) and associated production together with
top and bottom quarks ($gg \to H_1 t \bar t$ and $gg \to H_1 b \bar b$
respectively). As we saw for the branching ratios we again see
structures which are associated with the couplings of the Higgs boson
to various particles passing through zero. However, in contrast to the
earlier discussion, there are now three, rather than two, significant
values of $M_A$ where structure appears. The coupling of the Higgs
boson to a vector boson $V=W,\; Z$ with respect to the SM is given by
\begin{equation}
g^{\rm NMSSM}_{H_1VV}=O^H_{21}\; g^{\rm SM}_{H_1VV} \label{eq:ghvv},
\end{equation}
where $O^H_{21}$ is the same element appearing in
Eqs.(\ref{eq:ghuu}--\ref{eq:ghdd}), so when this mixing element
vanishes the vector boson fusion and Higgs-strahlung cross-sections
will disappear. The $H_1$ state at this point is {\it not} a purely
singlet state. The initial rotation of the doublet scalars by the
angle $\beta$ ensured that the only one of the doublet scalars has a
coupling to vector bosons. The vanishing of the $H_1VV$ coupling only
requires that there be none of this doublet state mixed in with $H_1$;
the $H_1$ field may (and does) still contain some of the doublet
scalar which does not couple to the vector bosons.  This $M_A$ point
where the $H_1VV$ coupling vanishes is very close to the point where
the $H_1t \bar t$ coupling vanishes because the first term on the
right-hand-side of Eq.(\ref{eq:ghuu}) is suppressed by $1/\tan \beta$.

For the lower values of $M_A$, where the Higgs decay to $b \bar b$ is
suppressed, this Higgs boson may be visible via its decay to $\gamma
\gamma$ (with a branching ratio $\gtrsim 0.1\%$ for $M_A \lesssim
480$~GeV). However, as the $\gamma \gamma$ branching ratio is turned
off at higher $M_A$, seeing this Higgs boson will become much more
challenging. Although the cross-section remains relatively large, the
Higgs boson almost always decays hadronically and the signal has a
very large QCD background. The only significant non-hadronic decay is
the Higgs decay to $\tau$-pairs with a branching fraction of
approximately $10\%$, but this also has large SM backgrounds. \\

The chosen scenario is extremely challenging for the LHC, but it is by
no means a ``worst-case scenario''. For example, increasing the value
of $\tan \beta$ would increase the separation between the $b$-quark
and vector boson switch-off points, moving the $M_A$ range with an
enhanced $H_1 \to \gamma \gamma$ branching ratio out of the allowed
region. Alternatively, increasing the value of $\kappa v_s$ slightly
would lead to a light Higgs boson sitting right on top of the
$Z$-peak, making it very difficult to disentangle from the SM
backgrounds. If the value of $\kappa v_s$ is significantly larger (and
$|A_{\kappa}|$ not too large), the singlet dominated scalar would be
heavy enough to decay to a vector boson pair, making its detection
much easier. However, if the value of $M_A$ is such that the coupling
of Eq.(\ref{eq:ghvv}) vanishes, these golden channels would be lost.

\subsection*{6~~~LC Production}

\begin{figure}[htb]
\begin{center}
\includegraphics[scale=0.6]{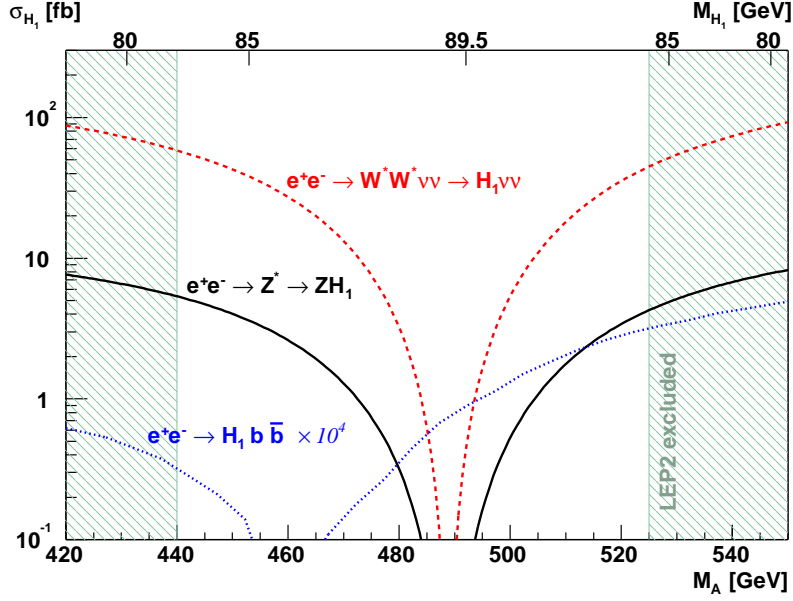}
\caption{\it 
Production cross-sections for the lightest scalar Higgs boson at a
$\sqrt{s}=500$~GeV LC, as a function of $M_A$ for $\lambda=0.3$,
$\kappa=0.1$, $\tan \beta= \tan \beta_s =3$ and $A_{\kappa}=-60$~GeV. 
The cross-section for $e^+e^- \to H_1 b \bar b$ has been multiplied 
by $10^4$.}
\label{fig:lc500}
\end{center}
\end{figure}

The vanishing of the $HVV$ couplings in the region of interest is
particularly significant for a LC since the most promising production
mechanisms are vector boson fusion, e.g.\ $e^+e^- \to W^+W^-\nu \bar \nu
\to H_1 \nu \bar  \nu$, and Higgs-strahlung, $e^+e^- \to Z^* \to
ZH_1$. The cross-sections for these processes at a $\sqrt{s}=500$~GeV
LC are plotted in Fig.(\ref{fig:lc500})
for our parameter choice, as a function of $M_A$, and show the
distinctive vanishing of the $H_1VV$ coupling.  Nevertheless, the
lightest scalar Higgs boson would be seen by these channels for all of
the $M_A$ range except for a small window around $490$~GeV. In
contrast to the LHC, for most of the the observable region decays to
$b \bar b$ and/or $\tau^+ \tau^-$ could be easily used due the LC's
relatively background free environment. For $M_A$ values where the
bottom and $\tau$ couplings vanish, the decays to $\gamma \gamma$ and
charm may be used instead. Indeed, as long as the Higgs-strahlung cross
sections are non-negligible, the associated Higgs particles can be
discovered irrespective of the Higgs decay properties.

It is difficult to see what production mechanism could be used to
close the remaining window around the critical point where the $HVV$
couplings vanish. Higgs production in association with a top quark
pair, $e^+e^- \to H_1 t \bar t$, is vanishingly small here because of
the proximity of the $H_1VV$ and $H_1 t \bar t$ ``turning-off'' points
(they will move even closer as $\tan \beta$ is increased). The
production in association with bottom quarks is shown in
Fig.(\ref{fig:lc500}), multiplied by a factor of $10^4$ to be visible
on the same scale. Generally, this production process has three
contributing sub-processes: Higgs-strahlung, $e^+e^- \to ZH_1$,
followed by the $Z$ decay to a bottom quark pair; Higgs pair
production, $e^+e^- \to H_1A_i$ ($i= 1,2$) followed by the
pseudoscalar decaying to bottom quarks; and bottom quark pair
production, $e^+e^- \to b \bar b$ followed by the radiation of $H_1$
off a bottom quark. The first contribution is very closely related to
the Higgs-strahlung already shown in Fig.(\ref{fig:lc500}) [simply
multiplied by the $Z \to b \bar b$ branching ratio], so contains no
new information and is {\it not} included in the $e^+e^- \to H_1 b
\bar b$ cross-section shown. The second contribution is only
kinematically allowed for the lightest pseudoscalar Higgs boson and is
vanishingly small because two small mixings are needed (neither scalar
nor pseudoscalar singlet fields have a $Z$ coupling). Therefore
the remaining process is dominated by Higgs radiation off bottom quarks,
and although this switches off at a different $M_A$ value, it is too
small to be useful because of the small bottom quark Yukawa coupling.

\begin{figure}[htb]
\begin{center}
\includegraphics[scale=0.6]{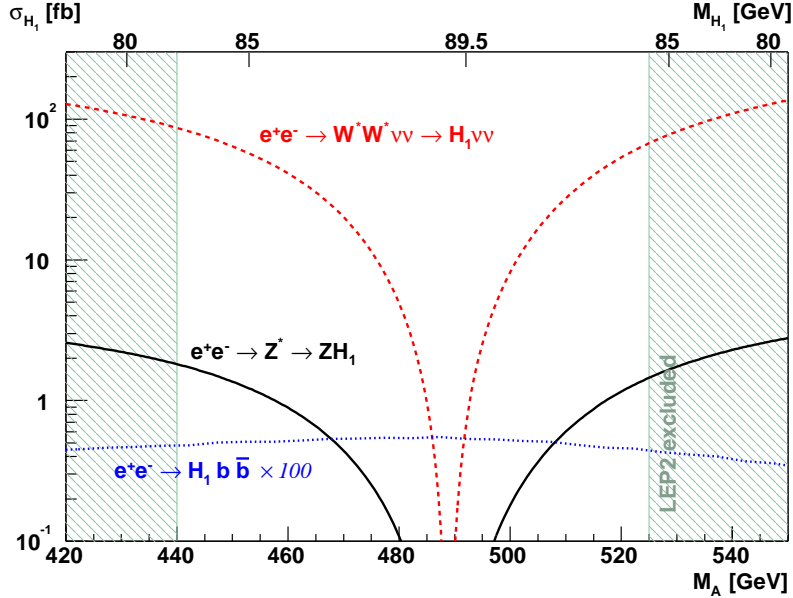}
\caption{\it 
Production cross-sections for the lightest scalar Higgs boson at a
$\sqrt{s}=800$~GeV LC, as a function of $M_A$ for $\lambda=0.3$,
$\kappa=0.1$, $\tan \beta= \tan \beta_s =3$ and $A_{\kappa}=-60$~GeV. 
The cross-section for $e^+e^- \to H_1 b \bar b$ has been multiplied 
by $100$.}
\label{fig:lc800}
\end{center}
\end{figure}

At a LC with $\sqrt{s}=800$~GeV, these cross-sections are modified as
shown in Fig.(\ref{fig:lc800}). 
The t-channel $W$-fusion cross-section increases, while the
$s$-channel Higgs-strahlung cross-section decreases, but the overall
$M_A$ dependence remains the same, with both cross-sections vanishing
at around $490$~GeV. The $e^+e^- \to H_1 b \bar b$ associated
production cross-section has increased dramatically due to the opening
up of $e^+e^- \to H_1A_2$, which was kinematically disallowed at
$\sqrt{s}=500$~GeV. Since this new contribution contains no $H_1 b
\bar b$ coupling, the cross-section no longer vanishes at around
$460$~GeV, but unfortunately it is still too small to be of practical
use\footnote{This cross-section has been calculated under the
assumption of a fixed width (of $1$~GeV) for $A_2$, and is only
intended to present an order of magnitude estimate.}. \\

Increasing $\kappa v_s$ and thus the singlet dominated masses only
reduces the production cross-sections in line with the expectations of
a reduced phase space. If the singlet dominated scalar is heavy
enough, and $M_A$ is far enough away from its critical value, the
scalar will decay to vector bosons, making its discovery easier.

\subsection*{7~~~Conclusions}

In this contribution we have considered a particularly challenging
NMSSM scenario, presenting masses, branching ratios and production
cross-sections at both the LHC and a future $e^+e^-$ LC. Such
scenarios have a Higgs spectrum very similar to the MSSM, i.e.\ nearly
degenerate heavy charged, scalar and pseudoscalar states and a light
Higgs boson at around $120$--$140$~GeV, supplemented by an additional
singlet dominated scalar and pseudoscalar.  We have seen that there is
still room allowed by LEP for the singlet dominated Higgs boson to be
very light, i.e.\ $\lesssim M_Z$. Despite having reasonably large
production cross-sections at the LHC, this light Higgs boson would be
difficult to see since its mainly hadronic decays cannot be easily
untangled from the SM backgrounds. At a LC, this light scalar can be
seen via vector boson fusion and Higgs-strahlung for most of the
parameter range, except for a small region where the Higgs-vector
boson coupling vanishes.  The observation of this light scalar state
at the LC, in addition to the light MSSM type scalar at the LHC,
provides an unambiguous signal for an extended supersymmetric theory
beyond the minimal version.  If this Higgs boson is discovered at a LC
but is missed at the LHC, LC input would be vital in providing
information for trigger and background removal when the LHC endeavours
to confirm the discovery.

We have also seen that a such a light Higgs boson may place
restrictions on the masses of the heavier Higgs bosons. For small
$\kappa v_s$, in order to avoid detection of the light scalar at LEP,
we require $M_A \approx \mu \tan \beta$. [The veracity of the
pre-condition ``small $\kappa v_s$'' may be ascertained by also
observing the singlet dominated pseudoscalar, by e.g.\ $e^+e^- \to t
\bar t A_1$, and making use of the approximate sum rule of
Eq.(\ref{eq:sumrule}).] This prediction for the heavy Higgs boson
masses would be invaluable to the LHC.

In this scenario the $H_2$, $H_3$ and $A_2$ will be present, looking
very much like the MSSM Higgs bosons $h$, $H$ and $A$ respectively
with slightly altered couplings and could be detected in the usual
way.

For heavier singlet dominated states, the position of the LHC is more
favourable, since the clean decay to vector bosons opens up [although
again, this is not useful over the entire $M_A$ range].  Also the
LHC's kinematic reach will prove useful in discovering or ruling out
very heavy singlet dominated Higgs states. On the other hand, if the
extra singlet dominated Higgs boson is found to be almost degenerate
with the lightest doublet dominated Higgs boson, LC precision may be
required to disentangle the two states.

In summary, in order to provide complete coverage over the NMSSM
parameter space, both the LHC and an $e^+e^-$ LC will be needed. Not
only can the LC probe areas where the LHC cannot, it can provide
valuable input to the LHC investigation of the NMSSM Higgs sector.

\subsection*{Acknowledgments}
DJM thanks the University of Southampton for their hospitality while
part of this work was carried out.

\end{document}